\documentclass[prb,twocolumn,showpacs,amsmath,amssymb]{revtex4}
\usepackage{graphicx}% Include figure files
\usepackage{dcolumn}% Align table columns on decimal point
\usepackage{bm}% bold math

\begin{document}

\title{How to detect gap nodes of a superconductor by 
angle-resolved specific heat experiment?} 

\author{P. Miranovi\' c}
\affiliation{Department of Physics, 
University of Montenegro, 81000 Podgorica, Serbia and Montenegro}
\author{M. Ichioka}
\author{K. Machida}
\affiliation{Department of Physics, 
Okayama University, Okayama 700-8530, Japan}
\author{N. Nakai}
\affiliation{Yukawa Institute for Theoretical Physics, Kyoto University,
Kyoto 606-8502, Japan}

\date{\today}

\begin{abstract}
The specific heat oscillation in the mixed state of type II
superconductors is studied theoretically when rotating
field within a plane containing gap minimum and maximum.
The calculations are performed microscopically by solving quasi-classical
Eilenberger equation for vortex lattices.
 The field dependence of the oscillation amplitude can discriminate between the nodal
and anisotropic gap with a finite minimum and 
the oscillation phase gives the gap minimum position on 
the Fermi surface. These also provide a way to separate out
the anisotropic behavior due to the Fermi velocity.
\end{abstract}

\pacs{74.25Bt, 74.25Op, 74.20Rp}

\maketitle

\section{Introduction}

There has been much attention focused on exotic superconductors,
including high T$_{\rm c}$ cuprates and heavy Fermion materials
in recent years.
In addition to the spin structure or parity of the Cooper pair,
the orbital function or the gap structure on the Fermi surface
is decisive to characterize its superconductivity. These studies are
expected to lead to a new pairing mechanism.
Even in conventional superconductors the energy gap 
can vary, depending on the position on the Fermi surface.
The degree of the anisotropy in the gap function is an important factor in understanding 
a superconductor in question. To distinguish a nodal superconductor 
from an anisotropic one with a finite minimum gap is of particular importance because
 the gap function can not change its sign in the latter while it can in the former. 
Also it is crucial to determine the maximum and minimum gap positions 
on the Fermi surface.

It is now widely recognized that the zero-energy density of states (ZEDOS)
sensitively reflects the gap structure, which is probed by a variety of experimental methods
such as specific heat, thermal conductivity or scanning tunneling spectroscopy.
This is particularly true for physical quantities in the mixed state of type II superconductors.
Induced vortices under an applied field carry a certain amount of ZEDOS around each
vortex core which depends on the gap structure\cite{volovik,hasegawa,ang}. 
The Sommerfeld coefficient $\gamma(B)$
as a function of magnetic induction $B$, which is nothing but ZEDOS induced by vortices, is
found to be one of physical quantities to reveal the gap topology.
In fact, there have been several $\gamma(B)$ 
experiments\cite{sonier,ramirez,junod,hedo,nohara,mirmelstein} on such as 
2H-NbSe$_2$, V$_3$Si, Nb$_3$Sn or CeRu$_2$.
We recently demonstrate that precise measurement $\gamma(B)$ 
at low field gives rise to indispensable information on the gap anisotropy\cite{nakai}.
In order to better characterize the gap structure it is urgent to
provide further ways to analyze experimental data. For example, anisotropic
behaviors in a superconductor could come from the two main sources.
One is the gap structure itself and the other is the Fermi velocity anisotropy due to band
structure. It is often the case that these two kinds of anisotropy are mixed up 
and difficult
to separate out individually, leading to an ambiguous conclusion as for the gap structure.
Thus we are required to devise some method to disentangle these two anisotropy effects.

Recently the angle-dependent specific heat experiments for the mixed state in
several superconductors LuNi$_2$B$_2$C \onlinecite{park}, CeCoIn$_5$ \onlinecite{aoki} 
and Sr$_2$RuO$_4$ \onlinecite{deguchi} have been
performed to yield characteristic oscillation pattern in 
$\gamma(B)$. A few percent oscillation amplitude relative to the total 
in these experiments is generally consistent with the theoretical estimate\cite{ang}
for nodal superconductors or strongly anisotropic
gap superconductors. However, it remains open to distinguish between them.
Specifically, it is reported that  the oscillation amplitude
becomes vanishing toward $B\rightarrow 0$ in Sr$_2$RuO$_4$ \onlinecite{deguchi} 
while it remains a finite value in LuNi$_2$B$_2$C \onlinecite{park} and 
CeCoIn$_5$ \onlinecite{aoki}. 
It was speculated that in the former
either the gap structure has a finite minimum gap or the gap in the 
minor band may mask the oscillation in lower fields. Thus we need 
to know precise behavior of the field-dependence of the oscillation amplitude for 
the two cases.

The purpose of the present paper is to examine the oscillation amplitude of the
angle-dependent Sommerfeld coefficient $\gamma(B)$ when $B$ rotates within 
a plane containing the gap minimum and maximum for several typical gap 
topologies, including line and point node superconductors and a 
superconductor with a finite minimum gap. We also study the oscillation behavior of 
$\gamma(B)$ for isotropic gap case with the anisotropic Fermi velocity.
It turns out that this anisotropy also yields a substantial specific heat 
oscillation under field rotation, but we will provide information on how to
distinguish it from the gap anisotropy case.
The existing data of the $\gamma(B)$ oscillation experiments
on LuNi$_2$B$_2$C, CeCoIn$_5$ and Sr$_2$RuO$_4$ are
analyzed in this regard. After a brief introduction to
quasi-classical framework to show how to calculate the
ZEDOS for various situations, we describe the results of the
ZEDOS oscillations to examine the
differences between node versus nodeless gap cases  in Section 3 and also
show the results for anisotropic Fermi
velocity in Section 4. The  Section 5 is devoted to the  point node case in
comparison with the line node
case to supplement the above analysis. In the Section 6 
we give a summary and discussions
on the existing data. All computations are done assuming a Fermi sphere.

\section{Quasi-classical theory and ZEDOS} 

The amplitude of the specific heat oscillations appears to be very small,
just few percents\cite{park,aoki,deguchi} as mentioned before. 
The smallness of the effect necessitates the use of
numerical solutions of Gorkov's microscopic equations of superconductivity 
to accurately estimate the specific heat amplitude. Quasi-classical approximation 
of Gorkov's equations, that we solve numerically here, is good as long as condition 
$k_F\xi\gg 1$ is met. Here, $k^{-1}_F$ is in the order of atomic length, 
and $\xi$ is coherence length.
Pairing interaction between electrons is modeled as  $V(\bm v,\bm v')=V_0\Omega(\bm v)\Omega(\bm v')$. This greatly simplifies the analysis, since pairing potential
in this model can be written as $\Delta(\bm r,\bm v)=\Omega(\bm v)\Psi(\bm r)$. 
The orbital part of the pairing potential, $\Omega(\bf v)$, we simply call: the gap function.
Eilenberger equations then read as ($\hbar=1$)

\begin{equation}
\left[2 \omega+{\bm v}{\bm\Pi}\right] f(\omega,{\bm r},{\bm v})
=2\Delta(\bm r,\bm v)g(\omega,{\bm r},{\bm v}),
\end{equation}
\begin{equation}
\left[2 \omega-{\bm v}{\bm\Pi}^*\right] f^\dagger(\omega,{\bm r},{\bm v})
=2\Delta^*(\bm r,\bm v)g(\omega,\bm r,\bm v).
\end{equation}
Here ${\bm \Pi}=\bm\nabla+(2\pi i/\Phi_0)\bm A$ is gauge invariant gradient,
$\bm A$ is vector potential, $\Phi_0$ is flux quantum, $\bm v=\bm v(\bm k_F)$ 
is Fermi 
velocity defined as $\bm v(\bm k)=\bm\nabla_{\bm k}E(\bm k)$, with $E(\bm k)$ 
being energy function of electrons in band; Fermi wave vector 
$\bm k_F$ can be found from the equation $E(\bm k)=E_F$;
 $\omega=\pi T(2n+1)$ with integer $n$ is Matsubara 
frequency,
$g$ and $f$ are normal and anomalous Green's function and 
$f^\dagger(\omega,\bm r,\bm v)=f^*(\omega,\bm r,-\bm v)$. 
Normalization condition for the Green's functions is $g^2+ff^\dagger=1$. 

The selfconsistency equations for 
the gap function and current density are:
\begin{equation}
\Delta(\bm r,\bm v)=2\pi N_0 T \sum\limits_{\omega>0}^{\omega_D}
\;\int\limits_{FS} d^2 k_F'\;V(\bm v,\bm v')\rho(\bm k_F')f,
\end{equation}
\begin{equation}
\bm j=4\pi i|e|N_0 T\sum\limits_{\omega>0}
\;\int\limits_{FS}d^2k_F\;\rho(\bm k_F)\bm v g.
\end{equation}
Here, $\int_{FS}$ is integral over the Fermi surface, 
$\omega_D$ is cut off frequency, 
$N_0$ is the total density of states for one 
spin at the Fermi surface in the normal state
\begin{equation}
N_0=\int\limits_{FS}{\frac{d^2k_F}{(2\pi)^3}}\;
\frac{1}{|\bm v|},
\end{equation}
and $\rho(\bm k_F)$ is the angle resolved density of states at the Fermi surface:
\begin{equation}
\rho(\bm k_F)=\frac{1}{(2\pi)^3 N_0}\;
\frac{1}{|\bm v|},
\end{equation}
normalized so that:
\begin{equation}
\int\limits_{FS} d^2k_F\rho(\bm k_F)=1.
\end{equation}
Density of states $N(\bm r, E)$ is defined as

\begin{equation}
N(\bm r, E)=N_0\int\limits_{FS} d^2 k_F\;{\rm Re}\;
g(i\omega=E+i\delta,\bm r,\bm v)\rho(\bm k_F)
\end{equation}

We are interested in low temperature zero-energy density of states (ZEDOS)
$N(\bm r,E=0)$. This is because in the limit of small temperatures, 
$T\longrightarrow0$, ratio of specific heat in superconducting state 
$C_s$ and normal state $C_n$ is given by
\begin{equation}
\lim\limits_{T\rightarrow0}\frac{C_s}{C_n}=
\frac{\overline{N(\bm r,E=0)}}{N_0}\;.
\end{equation}
Here $\overline{N(\bm r,E=0)}$ is spatially averaged ZEDOS in superconducting state.
In our calculation we set $T=T_{\rm c}/10$.
Since ZEDOS and specific heat are proportional at low temperatures we use
these two terms concurrently throughout the text. Numerical procedure
for solving quasi-classical equations of superconductivity is described
in Ref. \onlinecite{impurity}. 
Magnetic field is measured in units $\Phi_0/(2\pi R_0^2)$, where
$\Phi_0$ is flux quantum, and $R_0=0.882\;\xi_0$ ($\xi_0$ is BCS coherence length).

Angular dependence of ZEDOS is already studied numerically for some typical
cases of nodal gap function: 3D d-wave, polar state, and axial state\cite{ang}. 
Also, ZEDOS is studied by using the high field approximate solution of 
Eilenberger equations \cite{dahm,kus1,kus2,udu1,udu2}  
The focus in that study was on field dependence 
of ZEDOS for characteristic magnetic field directions. Just to recall, as one
rotates magnetic field from the gap node direction toward gap maximum direction, 
ZEDOS increases in low field, while it decreases in fields near $H_{c2}$. 
In other words, specific heat oscillation amplitude changes sign with increasing
field. In this paper we compare ZEDOS oscillation amplitude for nodal and nodeless
superconductors in the limit of low fields. Along with the nodal gap structures already 
studied in Ref. \onlinecite{ang}, we present also results for some other typical 
cases known in the literature.

\section{Nodal gap vs Nodeless gap}

In order to distinguish the nodal gap superconductor 
from the nodeless superconductor with a finite minimum 
gap, we examine the following model for the gap structure
\begin{equation}
\Omega(\varphi,\theta)=\Omega_0(a)\sqrt{1+a\cos{4\varphi}}
\end{equation}
with $\varphi$ polar angle and $\theta$ azimuthal angle in polar coordinates.
The parameter $a$ measures the degree of anisotropy. For $a=0$ the gap function
is isotropic, while for $a=1$ the gap function reduces to two-dimensional (2D)
version of $d_{x^2-y^2}$ -wave superconductivity. We choose the prefactor $\Omega_0(a)$ 
so that the average of $\Omega^2$ over the Fermi surface is unity, independent
of $a$. ZEDOS is calculated for fields rotating in the basal plane with $\theta=\pi/2$.
Only two field directions are of our interest, $B$ along the 
gap minimum ($\varphi=(2k+1)\pi/4$), and $B$ along the gap maximum ($\varphi=k\pi/2$).
Here, $k$ is integer. We performed the calculation for two anisotropy parameters:
$a=0.5$ and $a=1$. In both cases, low field ZEDOS is always minimum when the
magnetic field is oriented along the gap minimum (node) and vice versa.

\begin{figure}[b]
\includegraphics[angle=270,scale=0.35]{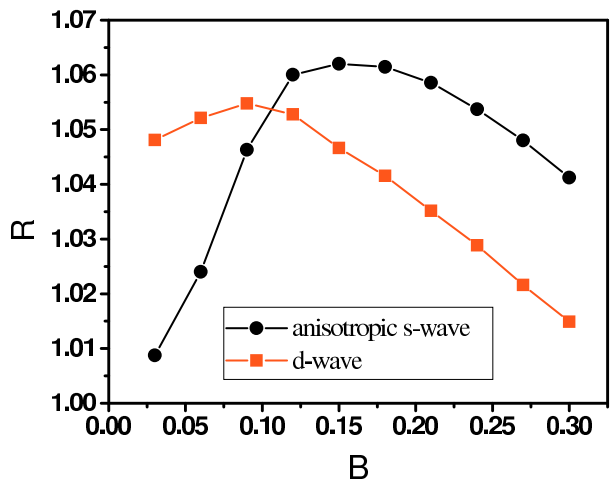}
\caption{Ratio $R=N(E=0,\varphi=0)/N(E=0,\varphi=\frac{\pi}{4})$
 of minimum and maximum ZEDOS for fields rotating in the basal
plane of the crystal. Full circles are for anisotropy parameter $a=0.5$, while
full squares are for 2D $d$-wave superconductor ($a=1$).}
\label{figsextended}
\end{figure}

In Fig. \ref{figsextended} the ratio $R=N(E=0,\varphi=0)/N(E=0,\varphi=\frac{\pi}{4})$ of 
maximum and minimum ZEDOS, for field rotating 
in the basal plane, is plotted for both anisotropy parameters. As is clear from 
Fig. \ref{figsextended}, there is a striking difference in low field dependence of
ratio $R$ for nodal $a=1$ and nodeless $a=0.5$ superconducting gap. 
Namely, for the nodal gap case ($a=1$) there is a finite amplitude ZEDOS
oscillation ($R\ne 1$) towards $B\rightarrow 0$.
This is contrasted with the nodeless gap case ($a=0.5$) where $R$ becomes unity at
lower fields, showing a maximum at the intermediate field region $B_{max}$. This means 
that the specific heat oscillation diminishes there. The field $B_{max}$ comes from
the physical reason that at $B<B_{max}$ the spatial extension of the ZEDOS is confined to
each vortex core region, yielding more or less isotropic ZEDOS landscape.
This ZEDOS feature does not cause the specific heat oscillation in this lower
field. The spatial extension of the ZEDOS landscape depends on the size
of the minimum gap because the gap acts as 
a potential for quasi-particles, that is, the zero-energy quasi-particles are strongly confined
and localized near each core. Thus $B_{max}$ signals the characteristic 
field where the localized zero-energy quasi-particles begin overlapping
and tends to become smaller as the minimum gap decreases.
Since in the nodal gap case the spatial extension of the ZEDOS is
extended to infinity, $B_{max}$ approaches zero, indicating that
$R$ stays a constant towards smaller fields.

\begin{figure}[h]
\includegraphics[angle=270,scale=0.35]{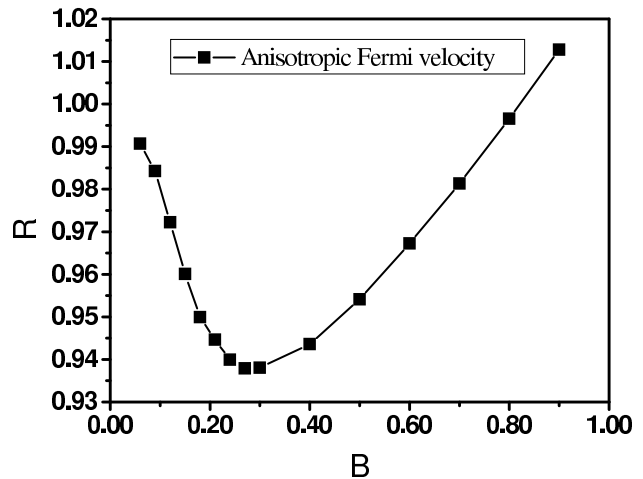}
\caption{Ratio $R=N(E=0,\varphi=0)/N(E=0,\varphi=\frac{\pi}{4})$
 of minimum and maximum ZEDOS in case of
anisotropic Fermi velocity modeled by Eq. (\ref{fermimodel}).}
\label{figfermi}
\end{figure}

\section{Anisotropic Fermi velocity}

We analyze the other type of anisotropy. The superconducting gap
is assumed to be isotropic, or to have the same value all over the Fermi surface, but the
Fermi velocity itself is anisotropic. The amplitude of the Fermi velocity 
on the Fermi sphere is
modeled as:
\begin{equation}
v(\varphi,\theta)=v_0(b)(1+b\cos4\varphi).
\label{fermimodel}
\end{equation}
The parameter $b$ measures the degree of anisotropy. For convenience
the prefactor $v_0(b)$ is chosen so that density of states in normal state
is independent of $b$. Four-fold variation of the Fermi velocity in the basal plane
is a quite simple 
model but will suffice for our purpose. Even in this hypothetical case
with  the isotropic gap and the anisotropic Fermi velocity, ZEDOS, and thus specific
heat, depend on magnetic field direction. In Fig. \ref{figfermi}, we plot
the ratio $R$ of minimum and maximum ZEDOS as a function of magnetic field.
The anisotropy parameter $b=0.5$ is used in the calculation. The ratio $R$ 
between maximum and minimum ZEDOS is of the order of a few percent. This is 
the same order 
of magnitude as in the above gap anisotropy cases. This is also the 
same order of magnitude observed \cite{park}in LuNi$_2$B$_2$C. This warns us that the effect of the Fermi 
surface structure on the specific heat oscillation cannot be neglected in 
intermediate fields. However, it is seen from Fig. 2 that extrapolation of maximum/minimum ratio 
$R$ to $B=0$ gives  $R=1$, i.e. disappearance of specific heat oscillation. 
This is qualitatively the same behavior as seen in the anisotropic gap case with a finite
minimum gap, shown in Fig. 1. Physically this is due to the spatial extension of the
zero-energy quasi-particle state: In the Fermi velocity anisotropy case
ZEDOS is confined to each core in a rather isotropic gap manner. Thus 
to induce the oscillation finite field is needed, above which because of the 
overlapping of the zero-energy quasi-particles, ZEDOS begins to exhibit an oscillation.

 It is also noticed that in the above $b>0$ case $R<1$, meaning that the four-fold oscillation
 pattern is maximally phase-shifted by $\pi/4$ from the gap anisotropy cases with $R>1$
 in Fig. \ref{figsextended}.   Needless to say, the sign $b$ is arbitrary for a given
 material, but it is physically plausible case $b>0$ when $a>0$ because in the angle-resolved
 DOS $N(\varphi)\propto 1/v(\varphi)$ the larger energy gap ($\varphi=0$) coincides with the larger angle-resolved
 DOS. This is indeed the case in boro-carbides.
 Therefore, we can clearly distinguish the two anisotropy effects by measuring the angle-dependent
 specific heat to monitor both the oscillation amplitude and its phase.

\section{Point node gap}

In this section we consider the point node gap structure.
Among several possible point node topologies we take up typical examples,
the so-called axial state and ``s+g''  model. The former is known to be realized
in superfluid $^3$He A phase and the latter is an candidate for boro-carbides\cite{maki}.
We also consider the polar state with a line node for comparison.

\begin{figure}[t]
\includegraphics[angle=270,scale=0.35]{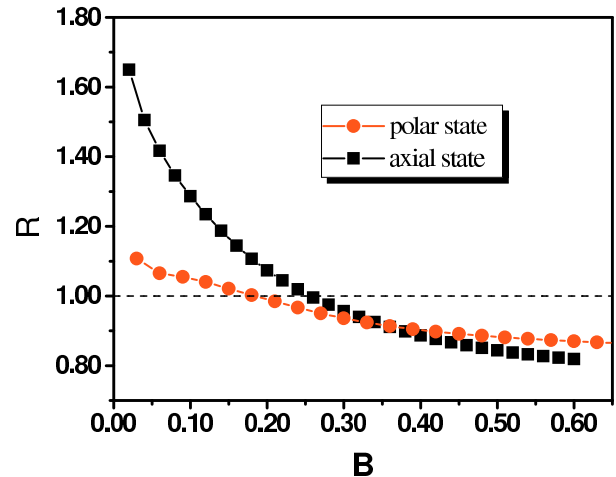}
\caption{Ratio $R=N(E=0,antinodal)/N(E=0,nodal)$ as a function magnetic induction
for polar and axial gap function.}
\label{figpolaraxial}
\end{figure}

\subsection{Axial and polar gap function}

The polar gap function has a horizontal line node in the crystal basal plane,
while the axial gap function is characterized by two point-like nodes
at the poles of the Fermi sphere. In polar coordinates with $\varphi$ and $\theta$ denote
polar and azimuthal angle, then the polar gap function is presented with
\begin{equation}
\Omega(\varphi,\theta)=\sqrt{3}\cos\theta,
\end{equation}
and the axial gap function is presented 
with 
\begin{equation}
\Omega(\varphi,\theta)=\sqrt{3/2}\sin\theta.
\end{equation}
The low temperature ZEDOS for 
these two gap functions is studied in detail in Ref. \onlinecite{ang}. By rotating 
magnetic field in a plane that contains $c$-axis, ZEDOS  periodically changes. The ZEDOS 
maximum appears for the field oriented along the gap node, and the ZEDOS minimum appears when magnetic 
field is directed along the gap maximum. Although field dependence of ZEDOS is already
presented in Ref. \onlinecite{ang}, for the purpose of this paper we present those 
data in a slightly different form.  Namely, we are interested in the amplitude of
ZEDOS oscillation as a function of $B$. Therefore, in Fig. \ref{figpolaraxial} 
the field dependence of ratio
$R=N(E=0,antinodal)/N(E=0,nodal)$ is shown. The low field ratio differs significantly
for the axial and polar gap functions. The important point to notice is that in both cases
ratio $R$, in the limit of low fields, differs from $1$. This means that there is
a finite amplitude of ZEDOS oscillation as $H\longrightarrow0$.
This is in accord with the conclusion in the previous sections.

\begin{figure}[t]
\includegraphics[angle=270,scale=0.35]{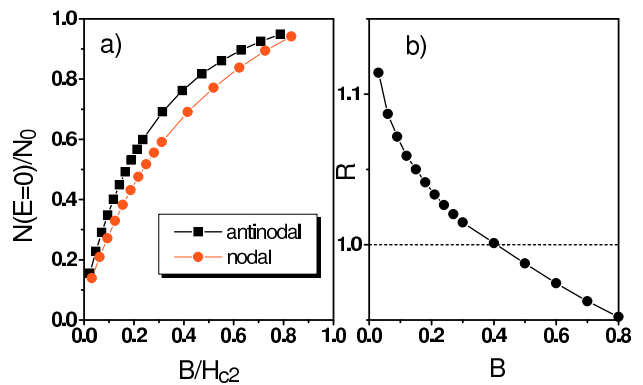}
\caption{a) Field dependence of ZEDOS for two field directions: nodal
and antinodal when fields are rotated in the basal plane. b) Ratio  $R=N(E=0,antinodal)/N(E=0,nodal)$ as a function magnetic 
induction.}
\label{figsplusg}
\end{figure}

\subsection{"s+g" model} 
  
It has been argued that the gap function in boro-carbides is a mixture of $s$-wave 
and $g$-wave superconductivity with aptly chosen weighting factors so that the pairing 
potential $\Omega(\varphi,\theta)$ does not change sign on the Fermi surface but has 
point-like zeros:
\begin{equation}
\Omega(\varphi,\theta)=\sqrt{\frac{315}{379}}\left(1-\sin^4\theta\cos4\varphi\right).
\end{equation} 
Like in previous examples of anisotropic pairing functions, we choose the Fermi surface as
a sphere to see the effect of the gap structure on low temperature thermodynamics.  
As we have already shown, even the anisotropy of Fermi surface alone can account for
direction dependent specific heat. The effect of Fermi surface disappears only in the 
limit of low fields. Having in mind that band structure in boro-carbides is far
from being isotropic, this simple "s+g" gap function on Fermi sphere, may not be
an appropriate model which can accurately estimate the amplitude of specific heat
oscillation in boro-carbides, but we can gain qualitative tendency in this case.    
In Fig. \ref{figsplusg}a) the field dependence of low temperature zero-energy DOS
is shown for nodal and antinodal field directions.  
We define the antinodal direction as a direction in the basal plane with 
maximum value of superconducting gap.
Magnetic induction $B$ is scaled 
with $H_{c2}{\rm(antinodal)}$ and $H_{c2}{\rm(nodal)}$ respectively. Note that $H_{c2}{\rm(antinodal)}>H_{c2}{\rm(nodal)}$.  As one may already anticipate, 
difference between the maximum ZEDOS 
(field along antinodal direction) and the minimum ZEDOS (field along nodal direction)
remains finite in low fields. Thus their ratio $N(E=0,antinodal)/N(E=0,nodal)\ne 1$.
This is shown in Fig \ref{figsplusg}b).
Noteworthy is comparison with gap structure of the axial state $\Omega(\varphi,\theta)=\sqrt{3/2}\sin\theta$. Both, the axial state and ``s+g" model,
have point like nodes, while their field dependence of ZEDOS is different.
The $\gamma(B)$ behavior as a function of $B$ 
is not governed only by the gap node topology, point-like or 
line-like nodes. It is not unique for all types of gap structures with
point-like (or line-like) nodes. It rather reflects the gap 
value on average, and it is rather sensitive to the functional form of the 
gap function in the vicinity of node. 

\section{Discussions on the specific heat experiments}
In this section we discuss the angle dependent specific heat
experiments on three materials, LuNi$_2$B$_2$C, CeCoIn$_5$
and Sr$_2$RuO$_4$ in the light of the present calculations.

\subsection{Boro-carbides}
Park {\it et al}cite{park}  measure the angle-dependent $\gamma(B)$ for LuNi$_2$B$_2$C and 
detect the four-fold oscillations in various fields, identifying the gap minimum
is located in [100] direction of tetragonal crystal because the oscillation maximum is 
in [110] direction. This identification is supported by our present result.
The oscillation amplitude becomes smaller as $B$ decreases.
However, it is rather difficult to judge whether the gap has a node or
a finite minimum gap from their experiment where the detailed
low field data are lacking. In connection with other experiments which suggest strong anisotropic
gap\cite{watanabe} or point node gap\cite{izawa-boro} in this system, 
it is interesting to extend their measurement to lower field to determine
the precise gap structure.

Izawa {\it et al} \cite{izawa-boro} show in their angle-dependent 
thermal conductivity measurement: (1) In the similar oscillation pattern
the maximum appears in [110], coinciding with that in Park {\it et al}.
(2) The oscillation amplitude diminishes when field rotates conically
out of the basal plane.
For the polar angle $\theta>\pi/4$ it almost vanishes. They conclude
a point node gap located along [100] direction. This assertion is based on a 
theoretical calculation\cite{maki} of angle dependence of thermal conductivity.
It may be informative in this connection to show our result:
We have also performed the calculation\cite{pedja} when field rotates conically
to check how the oscillation amplitude varies as a function of the polar angle
$\theta$. It is found that it decreases quickly as $\theta$ increases from
zero for both vertical line node and the ``s+g" point node cases, being unable to  distinguish
these two cases by conical field rotation of specific heat experiment.

\subsection{CeCoIn$_5$}
Aoki {\it et al} \cite{aoki} perform the angle \-dependent specific heat 
experiment in this system, observing a substantial oscillation 
amplitude when rotating field within the basal plane of tetragonal symmetry 
crystal. In the oscillation pattern the maximum occurs for [110] direction.
This suggests the d$_{xy}$ gap function because the oscillation amplitude
stays constant towards lowest fields. This conclusion appears to be inconsistent 
with the angle-dependent thermal conductivity experiment by Izawa {\it et al} \cite{izawa-115},
who conclude the d$_{x^2-y^2}$ gap function. Note, however, that their data themselves
exhibit the oscillation maximum for [110] direction, consistent with Aoki {\it et al} \cite{aoki}.

\subsection{Sr$_2$RuO$_4$}
According to Deguchi {\it et al} \cite{deguchi}, who measure the angle-dependent
specific heat on this system by rotating field within the basal plane 
of tetragonal crystal, the four-fold oscillation amplitude decreases below a threshold
field $\sim 0.3T$ and changes its sign near B$_{c2}\sim 1.5T$. 
The existence of the threshold field is 
in accord with our calculation where the gap structure has a finite minimum gap,
definitely excluding the vertical line node in the so-called main $\gamma$ band.
(We can not say anything about the possible horizontal line node\cite{yh}.)
An interesting point  in this system is the fact that we know accurately the
Fermi velocity anisotropy in the $\gamma$ band, where the larger Fermi velocity
is directed to [110]. Since the observed oscillation maximum occurs along [110], 
the experiment unambiguously excludes the oscillation due to the Fermi velocity 
anisotropy. Combining these two facts we conclude that there exists the 
anisotropic gap structure with a finite minimum  in the basal plane of
the main $\gamma$ band in Sr$_2$RuO$_4$. We can not commit 
ourselves on further 
conclusion concerning the minor $\alpha$ and $\beta$ bands or the horizontal
line node based on the existing data by Deguchi {\it et al} \cite{deguchi}.

\section{Summary and conclusion}

In this paper, we have calculated the zero-energy density of states
in the mixed state at low temperature by employing quasi-classical
Eilenberger formalism, which is valid for a wide variety of superconductors.
We have focused on the angle dependence of the zero-energy density of states,
which is directly measured through specific heat experiment as the
Sommerfeld coefficient, for superconductors with both nodal or nodeless 
gap structures.

We have demonstrated that the specific heat angular dependence
provides useful information concerning the gap structure, namely, the 
position of the node or the gap minimum on the Fermi surface
and also the existence or non-existence of the gap node.
Furthermore, we give information to distinguish two
sources of the anisotropy, either due to the gap itself or due to
the Fermi velocity of band structure. These proposed methods, 
we believe, add yet another dimension to firmly establish the gap
structure.

A few examples studied here does not exhaust all possibilities
for superconducting gap function. Neither all kind of different Fermi surface 
structures can be covered. Nevertheless, from the comparison of field 
dependence of ZEDOS in: a) nodal, and b) fully gaped anisotropic superconductors,
one can conjecture a behavior that is common for each group of superconductors.
In nodal superconductors the specific heat oscillation amplitude persists down to 
very low fields. In contrast to this behavior, fully gapped superconductors,
as one decreases magnetic field the oscillation amplitude gradually diminishes. 
To remind the reader again, only low temperature specific heat is discussed here.

\section{Acknowledgments}
We are grateful for useful discussions and communication
with Y. Matsuda, T. Sakakibara, K. Izawa, J. Sonier, K. Deguchi, Y. Maeno, A. Junod, and M. Salamon.

\end{document}